# Surface Aggregate Structure of Nonionic Surfactants on Silica Nanoparticles

**Dersy Lugo,**[a] **Julian Oberdisse,**[b] **Matthias Karg,**[a] **Ralf Schweins,**[c] **and Gerhard H. Findenegg**[*a]

[a] *Institut für Chemie, Stranski Laboratorium für Physikalische und Theoretische Chemie, Sekr. TC 7, Technische Universität Berlin, Strasse des 17. Juni 124, D-10623 Berlin, Germany.*
[b] *Laboratoire des Colloïdes, Verres et Nanomatériaux, UMR 5587 CNRS, Université Montpellier II, F-34095 Montpellier, France.*
[c] *Institut Laue-Langevin, DS / LSS group, B.P. 156, F-38042 Grenoble CEDEX 9, France.*

**Abstract**

The self-assembly of two nonionic surfactants, pentaethylene glycol monododecyl ether ($C_{12}E_5$) and n-dodecyl-β-maltoside (β-$C_{12}G_2$), in the presence of a purpose-synthesized silica sol of uniform particle size (diameter 16 nm) has been studied by adsorption measurements, dynamic light scattering and small-angle neutron scattering (SANS) using a $H_2O/D_2O$ mixture matching the silica, in order to highlight the structure of the surfactant aggregates. For $C_{12}E_5$ strong aggregative adsorption onto the silica beads, with a high plateau value of the adsorption isotherm above the *CMC* was found. SANS measurements were made at a series of loadings, from zero surfactant up to maximum surface coverage. It is found that the spherical core-shell model nicely reproduces the SANS data up to and including the local maximum at $q$ = 0.42 nm$^{-1}$ but not in the Porod region of high $q$, indicating that the surface area of the adsorbed surfactant is underestimated by the model of a uniform adsorbed layer. A satisfactory representation of the entire scattering profiles is obtained with the model of micelle-decorated silica beads, indicating that $C_{12}E_5$ is adsorbed as spherical micellar aggregates. This behaviour is attributed to the high surface curvature of the silica which prevents an effective packing of the hydrophobic chains of the amphiphile in a bilayer configuration. For the maltoside surfactant β-$C_{12}G_2$ very weak adsorption on the silica beads was found. The SANS profile indicates that this surfactant forms oblate ellipsoidal micelles in the silica dispersion, as in the absence of the silica beads.

## I. Introduction

The adsorption of nonionic surfactants at solid surfaces is of interest because of its importance to a great number of industrial and technological processes associated with colloidal stability and detergency. Adsorption of surfactants from aqueous solutions onto flat surfaces,[1-5] colloidal silica[2,6-8] and in porous silica[9-12] has been studied extensively in the past in view of its practical importance and rich bulk and interfacial behavior. A common point emerging from these studies is that adsorption at hydrophilic silica surfaces represents a surface aggregation process similar to micelle formation in the bulk solution. Surface aggregation may lead to discrete surface micelles or to fragmented or complete bilayers, depending on the surfactant composition and relative sizes of the head groups and the tails of the surfactant. However, information about the interfacial aggregation of surfactant molecules is still limited and it is not yet well understood.

Structural information about adsorbed surfactant layers at the interface of the aqueous phase against flat solid surfaces has been gained mainly by neutron reflectometry (NR)[1,2,4] and atomic force microscopy (AFM).[13-17] Direct evidence about surface aggregate structures of amphiphiles at flat solid surfaces has been gained by non-contact AFM. Specifically, it has become clear that many of the aggregate geometries which exist in bulk solution may also develop at the solid/solution interface, depending on the hydrophilic or hydrophobic nature of the surface.[17] These findings are consistent with calorimetric studies which show that at hydrophilic surfaces the enthalpy of adsorption of nonionic surfactants in the surface-aggregation regime is of similar magnitude as the enthalpy of micelle formation in the bulk solution.[18] NR yields the scattering-length density profile $\rho(z)$ which comprises information about the laterally averaged structure of the surfactant layer. In favourable cases, when using partially deuterated surfactants and different contrast scenarios, information about the arrangement of the amphiphile molecules in complete (saturated) mono- or bilayer can be obtained by this method. An attempt to model NR data of adsorbed surfactant films based on the existence of discrete surfactant aggregates of well-defined geometry was presented by Schulz et al.[19] As a more direct approach to resolve laterally structured surfactant layers by a non-invasive technique, Steitz et al.[20] used grazing-incidence small-angle neutron scattering (GISANS) to study surfactant films at a hydrophilic silicon wafer in the regime below saturation coverage. In that study the mean surface concentration of the chosen nonionic surfactant could be varied over a wide range simply by varying the sample temperature at a constant bulk concentration somewhat below the *CMC*. The GISANS results indicated the existence of transient surfactant aggregates without a preferred structure at half-coverage of the surface, indicating that the picture of distinct surface aggregates of characteristic size and separation that emerges from AFM studies of saturated surfactant layers cannot be generalized to the situation below complete surface coverage.

Small-angle neutron scattering (SANS)[6,8] and fluorescent probe techniques[9] have been used to characterize the nature of surfactant layers adsorbed at particulate and colloidal solids in aqueous dispersions. In their pioneering SANS work Cummins et al.[6] studied alkyl polyoxyethylene ether ($C_nE_m$) surfactants on Ludox HS and TM silica sols and demonstrated the effects of temperature, sol type, surfactant type and surfactant concentration on the adsorption at the colloidal silica particles. The adsorbed layer was modeled as a layer of uniform density, and hence the form factor was that of a sphere plus an outer shell. Later, one of the present authors[8,21] analysed SANS data of a technical-grade nonionic surfactant (Triton X-100) on silica nanoparticles on the basis of a model with a well-defined number of micellar aggregates adsorbed at the silica surface. It was found that this *micelle-decorated silica* model accounts in a quantitative manner for the experimental scattering profiles at maximal surface coverage of the silica beads.

The present work was aimed to gain a better understanding of aggregate structures of nonionic surfactants on colloidal particles at surface concentrations below the plateau of the adsorption isotherm. In this context it was of interest to compare the behavior of surfactants with a low and a high adsorption capacity. A longer-term goal of this work is to gain a better understanding of the effects of surface curvature on the nature and relative stability of the surface aggregates. In this paper we present a SANS study of the structure of adsorbed surfactant layers on colloidal silica over a wide range of surface concentrations of the surfactant. A silica sol of uniform particle diameter of ca. 16 nm was prepared and used to study the adsorption behavior of two different types of nonionic surfactants, viz., alkyl polyoxyethylene ether ($C_{12}E_5$) and alkyl maltoside ($\beta$-$C_{12}G_2$). The SANS data are analyzed in terms of the core-shell model and the micelle-decorated silica model.

## II. Experimental

### II.1 Materials

Pentaethylene glycol monododecyl ether, $C_{12}E_5$ (Fluka, purity ≥ 98%, density 0.963 g cm$^{-3}$ at 293 K), dodecyl-$\beta$-maltoside, $\beta$-$C_{12}G_2$ (Glycon, purity > 99.5%, density 1 g cm$^{-3}$), tetraethyl orthosilicate, TEOS (Fluka, purity ≥ 99.0%), ammonia (Sigma-Aldrich, A.C.S. reagent, 30-33% in water), ethanol, $C_2H_5OH$ (Berkel AHK, purity ≥ 99.9%), and $D_2O$ (Euriso-top, 99.9% isotope purity), were used without further purification. Normal water used in all experiments was purified by a Milli-Q 50 unit (Millipore, Billerica, USA) reaching a resistance of 18.2 MΩ and filtered by a 0.22 μm membrane to remove solid matter. A colloidal silica suspension, Ludox SM-30 (30 wt-% suspension in water), was supplied by Sigma-Aldrich. It was dialyzed with purified water for 2 weeks with daily exchange of water and filtered with a Millipore Steril Filter (0.8 μm). The resulting suspension contained 15 wt-% of silica. To preserve colloidal stability the pH was adjusted to pH 9 by the addition of 0.1 M NaOH. This Ludox stock suspension was stored in a refrigerator at 280 K.

### II.2 Preparation of the silica particles

Monodisperse silica nanoparticles of 16 nm diameter were prepared in two steps by the Stöber synthesis.[22] First a selective growth of Ludox particles was performed in basic conditions using TEOS as the silica source. 12.0 mL of the Ludox stock dispersion was added to a mixture of 240 mL ethanol, 60 mL water and 15.2 mL ammonia (30-33% in water) and was stirred in a PE bottle. 102 μL of TEOS were then added and stirring was continued for 24 h. After this growth step the Ludox particles had increased in size while the polydispersity had decreased. Particle growth was monitored by dynamic light scattering (DLS) and transmission electron microscopy (TEM). To reach the desired particle diameter the overgrown Ludox particles were dialyzed against Milli-Q water for 1 week with daily change of water. After dialysis the dispersion was concentrated by a factor 6 by solvent evaporation, using a rotary evaporator (313 K, 160 mbar), then filtered and the pH was adjusted to pH 9. The density and concentration of the resulting dispersion was determined gravimetrically.

The specific surface area of the silica was determined by nitrogen adsorption using a Micromeritics Gemini III 2375 Volumetric Surface Analyzer. For this purpose the silica dispersion was dried at 218 K for two days using a Freeze Dryer Alpha 2-4 LD/plus (Martin Christ), and then dehydrated at 393 K for 1 h under vacuum.

The hydrodynamic radius $R_H$ of the silica nanoparticles was determined by DLS. Correlation functions were recorded at different scattering angles between 45° and 135° using an ALV goniometer setup with a Nd:YAG-Laser as light source ($\lambda$ = 532 nm). The constant output power was 150 mW. All measurements were done at 298 K controlled by a thermostated toluene matching bath. The correlation functions were generated using an ALV-5000/E multiple-$\tau$ digital correlator and analyzed by inverse Laplace transformation (CONTIN[23]).

### II.3 Surfactant assembly on the silica particles

Samples with a low concentration silica sol (1-3 wt-%) were prepared at pH 9 to minimize complications by particle interaction and aggregation. Silica dispersions in normal water were used in surface tension and DLS measurements. Samples with seven different concentrations of $C_{12}E_5$ in a *C/CMC* range from 0.69 to 2.6 were prepared and equilibrated overnight before the measurements (*CMC* = 6·10$^{-5}$ M, *cf.* Section III.2). SANS measurements were made with silica sols in a $H_2O/D_2O$ mixture of scattering length density $\rho$ = 3.54x10$^{10}$ cm$^{-2}$ which matches that of silica, so that the scattering contrast arises only from the surfactant (*cf.* Section III.3). The silica content of the colloidal dispersions in $D_2O$ and the $H_2O/D_2O$ mixture was determined gravimetrically. Samples of the colloidal sol with different surfactant concentrations were prepared, corresponding to four different surface concentrations of $C_{12}E_5$ (¼$\Gamma_{mx}$, ½$\Gamma_{mx}$, ¾$\Gamma_{mx}$ and $\Gamma_{mx}$), and to two different surface concentrations for $C_{12}G_2$ (¾$\Gamma_{mx}$ and $\Gamma_{mx}$), where $\Gamma_{mx}$ is the plateau value of the surfactant adsorption isotherm as determined by the surface tension measurements.

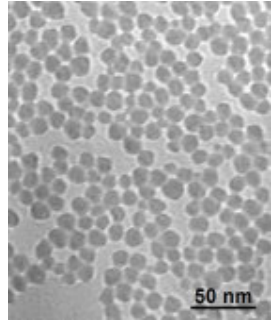

Fig. 1. Transmission electron micrograph of the silica used in this study.

The adsorption isotherm of $C_{12}E_5$ on the silica sol particles was determined by measuring the surfactant depletion in the supernatant solution after removal of the silica by centrifugation (4 h at 8500 rpm in a Universal 320R centrifuge). The equilibrium concentration in the supernatant was determined from its surface tension, using the $\gamma$-log(*C*) curve of aqueous solutions of $C_{12}E_5$ as a calibration curve.

The thickness of the adsorbed surfactant layer at the silica particles was estimated from the hydrodynamic radius $R_H$ of the silica nanoparticles in the absence and presence of $C_{12}E_5$, as described in Section II.2. A volume of 2 mL of each sample prepared for surface tension measurements was used in the DSL measurements. No corrections for the presence of micelles above the *CMC* were made. The viscosity of pure water was used in the calculation of $R_H$.

Small-angle neutron scattering experiments were performed on instrument D11 at ILL, Grenoble. The wavelength of neutron beam $\lambda$ was 0.6 nm with a spread $\Delta\lambda/\lambda$ of 9% fwhm. Data were collected at detector (64 x 64 pixels, each pixel 1 cm$^2$) distances of 1.2, 5 and 20 m from the sample, covering a range of the scattering vector *q* from 0.03 to 3 nm$^{-1}$. The samples were contained in standard 2-mm path length quartz cells, thermostated to ± 0.1 K using a Haake bath. Data reduction was done with the ILL standard software package according to ref. 24 and 25. Corrections were made for instrumental background, electronic noise, transmission, and empty cell. Incoherent background was determined with several H/D mixtures and interpolated for the desired concentrations. Normalization to absolute values of the scattered intensity (in cm$^{-1}$) was performed by measurements of the secondary calibration standard water (H$_2$O). The wavelength-dependent effective differential cross-section (d$\Sigma$/d$\Omega$)$_{H_2O}$ = 0.905 cm$^{-1}$ of H$_2$O for the $^3$He detector (at a wavelength of 0.6 nm) was taken from ref. 26. This is important for the present study, because it allows for an independent determination of the volume (or mass) of scattering objects. All theoretical curves are convoluted with the resolution function for each sample-to-detector distance. This includes the angular resolution due to the finite collimation, the detector cells, and the wavelength spread of the mechanical velocity selector.[27, 28]

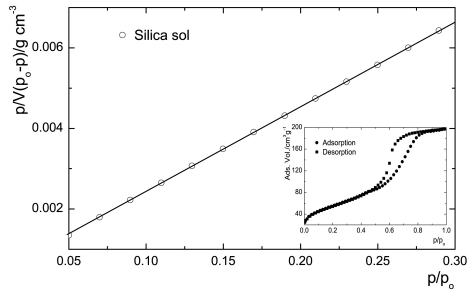

Fig. 2. BET plot of the nitrogen adsorption on the silica; the full adsorption isotherm is shown in the inset.

## III. Results

### III.1 Characterization of the silica particles

Electron micrographs show that the silica particles, prepared by Stöber synthesis using Ludox SM-30 as the starting material, are approximately spherical in shape (Figure 1), with an average particle diameter of about 16 nm. A somewhat larger diameter, 18 nm, was found by DLS. The aqueous dispersions at pH 9 had a zeta potential of -46 mV and a conductivity of about 120 μS cm$^{-1}$. The high zeta potential indicates a high surface charge of the particles which implies that the silica dispersion is electrically stabilized.

The nitrogen adsorption isotherm of the dried silica was analysed by the BET method in a range of relative pressures $p/p_0$ from 0.05 to 0.3 (Figure 2). A specific surface area as = 205 m$^2$g$^{-1}$ and BET constant $C_{BET}$ = 65 was obtained. The value of as is well compatible with the particle radius of 8.2 nm obtained by SANS, which leads to a geometric surface area $a_{geom}$ = 168 m$^2$g$^{-1}$. The ratio $a_s/a_{geom}$ = 1.2 can be attributed to surface roughness of the beads and indicates a low porosity of the beads. The features of the nitrogen adsorption isotherm at relative pressures $p/p_0 > 0.5$ (see inset in Fig. 2) can be attributed to pore condensation in the voids between close-packed silica beads in the dried sample and are thus of no significance for the present work.

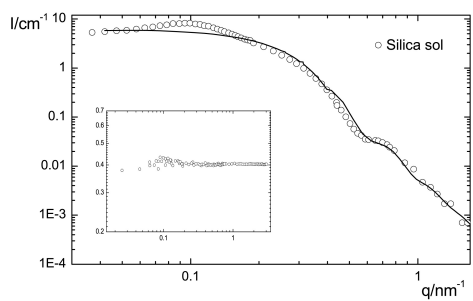

Fig 3. Scattering profile for 1.8 wt-% silica dispersion in nearly pure D$_2$O at pH 9 (298 K). The solid line represents the best fit with the log-normal size distribution function. The SANS profile showing mainly incoherent background for the silica dispersion in contrast matching H$_2$O/D$_2$O is shown in the inset.

The structure and shape of the silica particles was measured in a dilute silica suspension in D$_2$O-rich water ($\rho_D$ = 5.17·10$^{10}$ cm$^{-2}$). The scattering profile $I(q)$, shown in Figure 3, exhibits a pronounced oscillation near $q$ = 0.7 nm$^{-1}$, indicating a rather uniform size of the silica beads. The peak at $q_o \approx 0.1$ nm$^{-1}$ indicates repulsive electrostatic interactions between silica particles and it has been shown that this peak can be reproduced by standard theory.[8,29] Since the particle interactions are not of interest in the present context, we have not tried to describe this maximum at low $q$ quantitatively, but we have checked its consistency. Considering that the silica dispersion has a liquid-like structure, this structure peak at $q_o \approx 0.1$ nm$^{-1}$ can be used to estimate the silica particle radius, $R_{Si}$, by using a simple cubic lattice model (CLM), based on the conservation of silica volume on a unit cell, with the distance between the particles given *by* $D = 2\pi/q_o$. Since the volume of each particle can be estimated through $V = \varphi D^3$, where $\varphi$ is the volume fraction of the silica beads, $R_{Si}$ can be calculated by

$$R_{Si} = \sqrt[3]{\frac{6\pi\varphi}{q_o^3}} \qquad (1)$$

From the position of the peak we find $R_{si} = 8.1$ nm. If we concentrate on the form factor on the right of the $q = 0.7$ nm$^{-1}$ peak, the data can be described as a system of polydisperse spheres with a log-normal size distribution (*cf.* Appendix), with a polydispersity of 10 % and a mean particle radius of 8.2 nm. The relevant parameters of the colloid are summarized in Table 1.

**Table 1.** Scattering length density $\rho$, parameters of log-normal size distribution $R_{Si}$ and $\sigma$, average radius $\langle R_{Si}\rangle$, average surface area $\langle A \rangle$, average volume $\langle V \rangle$ of the silica particles

| $\rho$ (10$^{10}$ cm$^{-2}$) | $R_{Si}$ (nm) | $\sigma$ | $\langle R_{Si}\rangle$ (nm) | $\langle A \rangle$ (nm$^2$) | $\langle V \rangle$ (nm$^3$) |
|---|---|---|---|---|---|
| 3.54 | 8.20 | 0.10 | 8.24 | 8.62·10$^2$ | 2.42·10$^3$ |

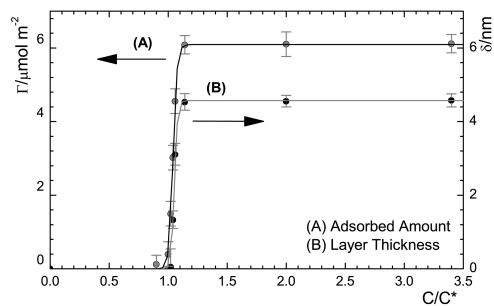

Fig. 4. Adsorption isotherm of C$_{12}$E$_5$ on the silica nanoparticles: (A) surface concentration $\Gamma$ as determined by surface tension measurements of the supernatant; (B) film thickness $\delta$ as obtained by dynamic light scattering (see text). The solid lines represent fits by the Gu-Zhu equation.

### III.2 Adsorption and film thickness of C$_{12}$E$_5$ on the silica

The adsorption isotherm of C$_{12}$E$_5$ on the colloidal silica, shown as curve (A) in Figure 4, exhibits a pronounced sigmoidal shape. The surface concentration $\Gamma$ remains very low up to a concentration $C^*$ at which $\Gamma$ increases sharply to a plateau value $\Gamma_{mx}$. The onset concentration is $C^* = 5 \cdot 10^{-5}$ M and the plateau value is reached at $C \approx CMC = 6 \cdot 10^{-5}$ M. The plateau value of the surface concentration, $\Gamma_{mx} = 6.10$ µmol m$^{-2}$, corresponds to 5.3·10$^{-21}$ mol C$_{12}$E$_5$ adsorbed per silica particle. The sharp rise in adsorption prior to the *CMC* indicates that surface aggregation caused by hydrophobic interactions between surfactant monomers comes into play, as was shown by fluorescence spectroscopy measurements.[9] Above the bulk *CMC*, the surfactant activity remains nearly constant and thus no further adsorption takes place. The shape of the adsorption isotherm is consistent with many literature reports and the plateau value of $\Gamma$ agrees with the value for C$_{12}$E$_5$ on silica wafers ($\Gamma_{mx} = 6.0$ µmol m$^{-2}$) obtained by ellipsometry by Tiberg et al.[3] A higher value ($\Gamma_{mx} = 7.5$ µmol m$^{-2}$) was reported for the adsorption on TK900 silica by Gellan and Rochester.[30]

The mean thickness $\delta$ of the adsorbed layer of C$_{12}$E$_5$ on the silica beads, shown as curve (B) in Figure 4, exhibits the same dependence on the solution concentration as the surface concentration $\Gamma$. The plateau value of the film thickness, $\delta_{mx} = 4.6$ nm, is higher than other values for C$_{12}$E$_5$ on silicon oxide surfaces reported in the literature, viz., 4.2 nm on oxidized silicon wafers obtained by ellipsometry,[3] 4.4 nm on TK900 silica obtained by X-ray diffraction,[30] or 4.5 nm on Ludox TM obtained by SANS by Cummins et al.[6] The somewhat higher value of $\delta_{mx}$ obtained by DLS in this work may be due to differences in $R_H$ between the bare and surfactant-coated silica particles caused by different hydrodynamic interactions of the two types of surfaces with the aqueous solution.

**Table 2.** Parameters of the Gu-Zhu equation (eq. 2), for the surface concentration ($\Gamma$) and film thickness ($\delta$) isotherms of $C_{12}E_5$ on the silica sol (298 K, pH 9)

|  | $\Gamma_{mx}$ (µmol m$^{-2}$) | $\delta_{mx}$ (nm) | $K$ | $C^*$ (mol L$^{-1}$) | $n$ |
|---|---|---|---|---|---|
| $\Gamma$ isotherm | 6.1 | - | 8.3·10$^{-2}$ | 5·10$^{-5}$ | 63 |
| $\delta$ isotherm | - | 4.6 | 8.4·10$^{-2}$ | 5·10$^{-5}$ | 45 |

The data for the surface concentration and film thickness can be represented by the mass action model of one-step formation of surface aggregates proposed by Gu and Zhu,[31]

$$Z = \frac{Z_{mx} K (C/C^*)^n}{1 + K(C/C^*)^n} \quad (2)$$

where $Z$ is either the surface concentration $\Gamma$ or the thickness of the adsorbed surfactant layer $\delta$ at equilibrium concentration $C$, $Z_{mx}$ is the limiting values of the surface concentration, $\Gamma_{mx}$, or of the adsorbed surfactant layer thickness, $\delta_{mx}$, at high $C$, $C^*$ is the onset concentration of aggregative adsorption, $K$ is the adsorption constant in the low-affinity region, and $n$ is the average aggregation number of the surface aggregates. Best-fit parameters of Eq. (2) are given in Table 2.

### III.3 Characterization of the adsorbed layer by SANS

SANS measurements were made to elucidate the structure of the adsorbed layer of the surfactant on the silica particles. A low silica concentration was chosen to suppress the influence of interparticle interactions. The silica concentration in the dispersion was fixed to 1.8 wt-% (0.90 vol.-%) in a H$_2$O/D$_2$O mixture that matches the scattering length density of the silica. The contrast-match point was determined experimentally; the quality of contrast-match is illustrated by the inset in Fig. 3. Figure 5a shows scattering profiles of silica dispersions containing $\beta$-C$_{12}$G$_2$ and C$_{12}$E$_5$ under contrast match for the silica particles. In both cases the surfactant concentration was by more than two orders of magnitude higher than the *CMC*. In the case of C$_{12}$E$_5$, the SANS profile exhibits a pronounced oscillation with a maximum at $q = 0.42$ nm$^{-1}$, indicating the existence of a surfactant layer on the surface of the particles.[6,8] On the other hand, no such features at intermediate $q$ are observed in the case of $\beta$-C$_{12}$G$_2$, indicating that this surfactant does not form an adsorbed layer at the surface of the silica particles, in agreement with reports in the literature.[5] Instead, $\beta$-C$_{12}$G$_2$ will form free micelles in solution and one expects that the scattering profile will resemble that of micellar solutions in the absence of silica. Scattering profiles of $\beta$-C$_{12}$G$_2$ and C$_{12}$E$_5$ in D$_2$O are displayed in Figure 5b.[32] A comparison of Fig. 5a and 5b shows that the scattering profiles of $\beta$-C$_{12}$G$_2$ in the presence and absence of silica are indeed similar. The experimental data for $\beta$-C$_{12}$G$_2$ in the silica dispersion could be fitted by an oblate ellipsoid core-shell form factor model characterized by an ellipsoid core with rotational semi-axis $a$ and equatorial semi-axis $b$, surrounded by a head group layer of constant thickness $d$ (*i.e.*, assumed to be the same along the short and long axes). The results are presented in Table 3. For $\beta$-C$_{12}$G$_2$ in the silica dispersion the best fit was obtained with $d$ values in the range 0.70 – 0.93 nm, in agreement with the values for the surfactant micelles in the absence of the silica ($d$ = 0.70-0.98 nm). Details of this analysis will be presented elsewhere.[32]

Figure 6 shows scattering profiles of silica dispersions with different amounts of adsorbed C$_{12}$E$_5$, corresponding to four different surface concentrations along the adsorption isotherm (¼$\Gamma_{mx}$, ½$\Gamma_{mx}$, ¾$\Gamma_{mx}$ and $\Gamma_{mx}$). As can be seen, both the scattered intensity and the oscillation in $I(q)$ at intermediate $q$ values increases with increasing surface concentration of C$_{12}$E$_5$. In particular, a pronounced maximum at $q = 0.42$ nm$^{-1}$ develops as $\Gamma$ increases. Below, we analyse these scattering profiles in terms of two form factor models.

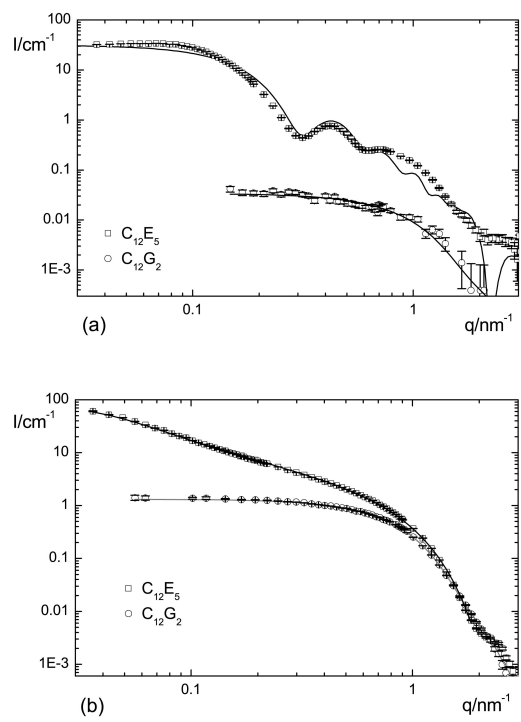

Fig. 5. SANS profiles of $C_{12}E_5$ and $\beta$-$C_{12}G_2$ (pH 9, 298 K): (a) in the presence of 1.8 wt-% colloidal silica in contrast matching $H_2O/D_2O$; surfactant concentrations (vol.-%): 0.8 ($C_{12}E_5$) and 0.1 ($\beta$-$C_{12}G_2$); solid lines represent fits by the micelle-decorated silica model ($C_{12}E_5$) or the ellipsoidal core-shell model for micelles ($\beta$-$C_{12}G_2$). (b) surfactant micelles (1 vol.-%) in $D_2O$ in the absence of silica (1 vol-% for each surfactant). Solid lines represent fits with the wormlike polymer model ($C_{12}E_5$) or the ellipsoidal core-shell model ($\beta$-$C_{12}G_2$).

**Table 3.** Parameters of the oblate-ellipsoid core-shell model for the SANS profile of the micelles of $\beta$-$C_{12}G_2$ in the silica dispersion in contrast-matching $H_2O/D_2O$

| Surfactant | $a$ (nm) | $b$ (nm) | $d$ (nm) | $n_{agg}$ |
|---|---|---|---|---|
| $\beta$-$C_{12}G_2$ | 1.41 | 2.84 | 0.93 | 136 |

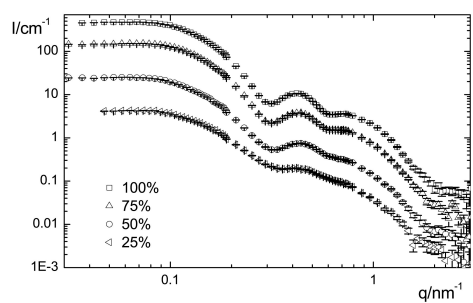

Fig. 6. SANS profiles of 1.8 wt-% silica in contrast-matching $H_2O/D_2O$ (pH 9, 298 K) in the presence of different amounts of $C_{12}E_5$, corresponding to surface concentration $\Gamma_{mx}$, ¾$\Gamma_{mx}$, ½$\Gamma_{mx}$ and ¼$\Gamma_{mx}$. The curves for higher surface concentrations are shifted vertically relative to that of ¼$\Gamma_{mx}$ by factors of 2 (50%), 6 (75%) and 14 (100%).

**III.3.1 Core-shell model**

As a first approach, a spherical core-shell model was adopted to fit the scattering profiles of the 1.8 wt-% colloidal silica with adsorbed $C_{12}E_5$ (Figure 7A). The particle form factor of this model, given by Eq. (A5) in the Appendix, has been used by Cummins et al.[6] in their study of surfactant adsorbed layers on colloidal silica. Figure 8a shows fits of the data at saturation surface coverage ($\Gamma_{mx}$) of $C_{12}E_5$ with three values of the layer thickness, viz. $\delta$ = 4.0, 4.4, and 4.6 nm. Clearly these values are causing too high scattering intensities except for the high-$q$ region. However, the maximum at $q = 0.42$ nm$^{-1}$ and the scattering intensities at lower $q$ are nicely reproduced when the intensities resulting from the model with $\delta$ = 4.0, 4.4, and 4.6 nm are multiplied by factors $f$ = 0.4, 0.34 and

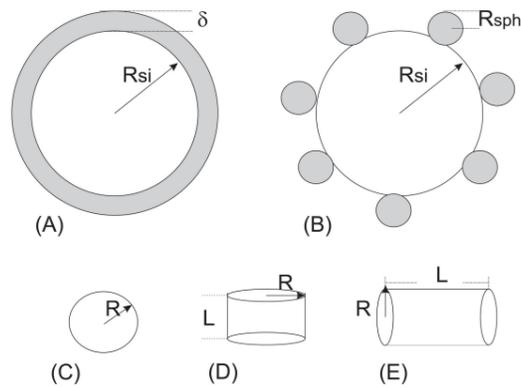

Fig. 7. Cartoons of surfactant self-assembly structures on silica beads: (A) spherical core-shell model; (B) micelle-decorated silica model with spherical surface micelles. Different forms of surface micelles: spheres (C); patch (D), and wormlike (E).

0.28, respectively, as seen in Fig. 8b. Similar results were obtained for surface concentrations ¾$\Gamma_{mx}$ and ½$\Gamma_{mx}$ with assumed values of the layer thickness of 3.6 and 2.8 nm, respectively. For the surface concentration ¼$\Gamma_{mx}$ the maximum in the scattering curve at $q = 0.42$ nm$^{-1}$ is rather weak and not well-reproduced with any value of the layer thickness (results not shown). However, in all cases the core-shell model curves fail to fit the high-$q$ regions of the scattering profiles when multiplied by any factor $f < 1$. Figure 8b indicates that the core-shell model adjusted to the maximum at $q = 0.42$ nm$^{-1}$ also fits the behaviour at lower $q$ but strongly underestimates the surface area of the total adsorbed surfactant (behaviour at high $q$). The large difference between core-shell model and experiment at large q (see Fig. 8b) cannot be removed by changes in the subtracted incoherent background within the limits of experimental uncertainty. Since the background is known to ±0.003 cm-1 in the present study, errors here are never sufficient to catch up the differences between the models at high-q (cf. Fig. 8). The effect of background subtraction on the data analysis in terms of the present models is demonstrated in the Supplementary Material. Hence the differences between core-shell model and experiment at high q imply that the surfactant is not forming a uniform layer but a different type of aggregate, e.g., surface micelles, which have a higher surface area at a given total adsorbed volume. Surfactant surface areas and volumes and numbers of adsorbed aggregates were estimated for different types of surface aggregates. It was found that the model of micelle-decorated silica beads,[8,21] with spherical surface aggregates as sketched in Figure 7B, gives a good representation of the data.

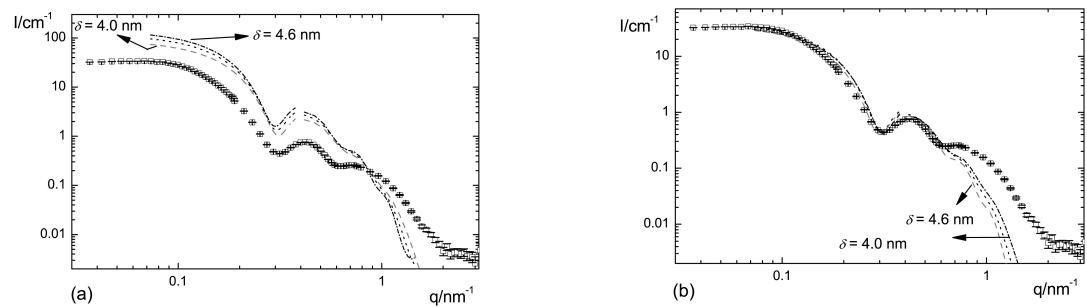

Fig. 8. SANS profile for surface concentration $\Gamma_{mx}$ of $C_{12}E_5$ on silica in contrast matching $H_2O/D_2O$: (a) Experimental data and intensities predicted by the spherical core-shell model with shell thicknesses $\delta$ = 4.0, 4.4 and 4.6 nm; (b) same as in (a) but predicted intensities multiplied by factors $f$ = 0.4, 0.34 and 0.28, respectively.

### III.3.2 Micelle-Decorated Silica Model

In applying the model of micelle-decorated colloidal silica beads the first step was to determine how much adsorbed surfactant volume is in excess ($I$, $q \rightarrow 0$), and how much surface is missing in the core-shell model ($I$, $q \rightarrow \infty$). The real adsorbed surfactant volume was estimated by introducing the effective core-shell volume fraction of the surfactant, $X = \sqrt{f}$, based on a fixed layer thickness of 4.0 nm. For example, at 100% surface coverage the intensities estimated by the core-shell model had to be multiplied by a factor $f = 0.40$ to match the measured intensities at low and intermediate $q$ values, which implies that $X = 0.63$, i.e. only 63% of the layer volume is occupied by $C_{12}E_5$. The total surface area of the adsorbed surfactant, $A_{total}$, was calculated from the specific surface area $S/V$ as determined by Porod's law ($S/V = A/2\pi\Delta\rho^2$, where $A$ is the Porod constant) and from the number density of silica particles, $N = 3.72 \cdot 10^{-6}$ nm$^{-3}$, by the relation $A_{total} = (S/V)/N$. The fitting of the scattering curves with this model was based on the film thickness as obtained in the spherical core-shell model, which reproduces very well the maximum at intermediate $q$. As seen in Figure 8, the predicted intensities for the core-shell model with layer thicknesses $\delta$ of 4.0, 4.4 and 4.6 nm are very similar. We have calculated the surface area and volume of the adsorbed surfactant based on these three values of $\delta$, and with these two new parameters it was possible to estimate the number and dimensional parameters of surface aggregates of different geometry. Calculations were performed for individual spherical, patchlike, and wormlike micelles (see Fig 7C-7E). They were made for an assumed layer thickness of 4.0 nm and full surface coverage ($\Gamma_{mx}$). The resulting geometrical parameters (radius $R$, length or height $L$) are given in Table 4. It was found that spherical aggregates adsorbed as separate entities represent the most acceptable geometry of the surface aggregates. The samples with surface concentrations ¼$\Gamma_{mx}$, ½$\Gamma_{mx}$, and ¾$\Gamma_{mx}$ conform to a similar behaviour. The respective model parameter are summarized in Table 5. The maximum in the scattering curves at low $q$ which is due to silica-silica interactions is not accounted for in the model in its present form but could in principle be included by a structure factor, as proposed by Despert et al.[8] The position of this interaction peak at 0.09 nm$^{-1}$ does not change due to the constant silica volume fraction used in our experiments. The position and amplitude of the maximum at $q = 0.42$ nm$^{-1}$ is very well reproduced by the model without any arbitrary factor, but some discrepancies between the experimental and predicted $I(q)$ appear in the high-$q$ Porod regime at each surface coverage, where the model overestimates (¼$\Gamma_{mx}$, ½$\Gamma_{mx}$) or underestimates (¾$\Gamma_{mx}$, $\Gamma_{mx}$) the surface area of the adsorbed surfactant aggregates, as seen in Figure 9. The quality of the fits was not significantly improved when in the former case (¼$\Gamma_{mx}$, ½$\Gamma_{mx}$) the radius of the micelles was increased at fixed number of adsorbed aggregates ($N_{sph} = 28$ for ¼$\Gamma_{mx}$, or $N_{sph} = 52$ for ½$\Gamma_{mx}$), nor in the latter case (¾$\Gamma_{mx}$, $\Gamma_{mx}$) when the radius of the aggregates was decreased at fixed $N_{sph}$. Such discrepancies may be due to the low surface coverage, possibly leading to the coexistence of populations of aggregates or unimers, which is not captured by our model because the adsorbed surfactant micelles have been modeled as monodisperse spheres, which induce strong oscillations in the high-$q$ region.

**Table 4.** Parameters of the micelle-decorated silica model for different geometries of the micellar surface aggregates of $C_{12}E_5$ at 100% surface coverage.

| geometry   | $R$(nm) | $L$(nm) | $N_{agg}$ |
|------------|---------|---------|-----------|
| spherical  | 2.0     | -       | 98        |
| patchlike  | 1.7     | -       | 93        |
| wormlike   | 2.0     | 2.9     | 114       |

**Table 5.** Parameters of the micelle-decorated silica model for spherical surface aggregates of $C_{12}E_5$ on silica particles: $X$, effective volume fraction of surfactant, $R_{sph}$, $N_{sph}$, radius and number of surface aggregates per particle.

| surf. concentration | $X$  | $R_{sph}$(nm) | $N_{sph}$ |
|---------------------|------|---------------|-----------|
| ¼ $\Gamma_{mx}$     | 0.23 | 2.2           | 28        |
| ½ $\Gamma_{mx}$     | 0.39 | 2.1           | 52        |
| ¾ $\Gamma_{mx}$     | 0.55 | 2.1           | 73        |
| $\Gamma_{mx}$       | 0.63 | 2.0           | 98        |

## IV. Discussion

The present study builds on the early work of Cummins et al.[6] who indicated that the SANS scattering profiles of $C_nE_m$ surfactants on Ludox silica could result from surface micelles, and on the first detailed description of the geometry of such surface micelles by one of the present authors.[21] The experiments were made with the home-made silica instead of commercial Ludox in order to attain a better control of the mean particle size, a lower polydispersity and a better colloidal stability of the samples. The surface chemistry of these samples may be somewhat different from that of Ludox, but this point is considered not important. A prominent feature of the scattering profiles of silica with adsorbed $C_{12}E_5$ is the maximum in $I(q)$ at $q = 0.42$ nm$^{-1}$ which can be taken as a measure of the amount of surfactant forming the adsorbed layer. The analysis of the scattering profiles has shown, however, that the spherical core-shell model adjusted to this maximum in $I(q)$ strongly underestimates the surface area of the adsorbed surfactant as extracted from the high-$q$ regime of the scattering profiles. This result suggests that the adsorbed surfactant is not forming a uniform bilayer but a different type of aggregate. On the basis of the micelle-decorated silica model, assuming spherical micellar surface aggregates, it is found that the number of adsorbed aggregates $N_{sph}$ increases as the concentration of surfactant in the system is increased (Table 5 and Fig. 9). The model also predicts that the micelle radius $R_{sph}$ somewhat decreases as the surfactant loading is increased. However, this predicted trend is probably not significant due to the limited resolution of the experimental data. The resulting mean value of $R_{sph}$ is similar to the cross-sectional radius of the wormlike micelles of $C_{12}E_5$ formed in aqueous solutions ($R \approx 2.4$ nm) as obtained from Guinier expression for cylindrical micelles.

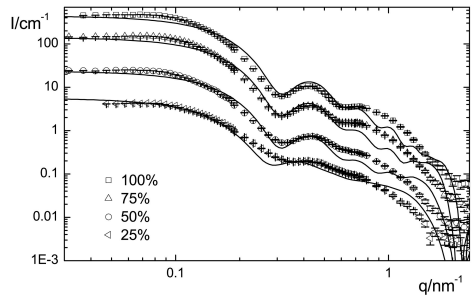

Fig. 9. SANS profiles for surface concentrations $\Gamma_{mx}$, $\frac{3}{4}\Gamma_{mx}$, $\frac{1}{2}\Gamma_{mx}$ and $\frac{1}{4}\Gamma_{mx}$ of $C_{12}E_5$ on silica in contrast matching $H_2O/D_2O$ (same data as in Fig. 6), and fits by the micelle-decorated silica model (full curves). Results for higher surface concentrations are shifted relative to those for $\frac{1}{4}\Gamma_{mx}$ by factors of 2 ($\frac{1}{2}\Gamma_{mx}$), 6 ($\frac{3}{4}\Gamma_{mx}$), and 14 ($\Gamma_{mx}$).

For spherical surface aggregates the micelle-decorated silica model predicts that the number of surface aggregates per silica particle, $N_{sph}$, can be as high as about one hundred (Table 5). This large number corresponds to a dense layer of surface micelles, as can be seen from graphical representations of model results (not shown), or by a simple estimation of the maximum number of spheres of radius $R_{sph}$ in contact with a central sphere of radius $R_{Si}$. Assuming that the smaller spheres arrange on a hexagonal lattice at a distance $(R_{Si} + R_{sph})$ from the centre, then $N^{mx} = (2\pi/\sqrt{3})(s + 1)^2$, where $s = R_{Si}/R_{sph}$. With $R_{Si} = 8.2$ nm and $R_{sph} = 2.0$ nm we find $N^{max} = 94$. Hence at the highest surface concentration, the layer geometry consistent with the scattering data is a dense layer of micelles. Admittedly, the exact geometry of this layer may differ from the one proposed here, *e.g.* the micelles might form bridges and build an undulating bilayer. In this case, however, high bending energy of the surfactant monolayer, forcing it to deviate from its preferred globular arrangement, would have to be accounted for. In summary, the micelle-decorated silica bead model accounts simultaneously for the high coverage, and the high specific surface area as measured (model-free) in the high-$q$ range.

The finding that $C_{12}E_5$ is adsorbed in the form of small surface micelles at the silica beads is remarkable in view of the fact that aggregates of smaller mean curvature (*e.g.*, wormlike micelles) are favoured in aqueous solutions and bilayer films are formed in the adsorption onto flat surfaces. We conjecture that the preference for small surface aggregates is a consequence of the high surface curvature of the silica nanoparticles which prevents an effective packing of the surfactant molecules in a bilayer film. Note that for particles of radius 8 nm and a bilayer thickness of 4 nm the mean area per molecule at the mid-plane of the bilayer is 50% greater than the respective area directly at the surface. Accordingly, an effective packing of the hydrophobic tails is not possible in a bilayer configuration (even if more molecules are accommodated in the outer layer than in the inner layer). Instead, small (roughly spherical) aggregates will be favoured, as these aggregates allow the most effective packing of the hydrophobic tails on small spherical particles. This argument applies not only to the region of low surface concentrations but also to the plateau region of the isotherm. Further work with silica particles of different size and surfactants of different packing parameters is needed to corroborate these findings. Such measurements are planned in our laboratories.

## V. Conclusions

We have used SANS to determine the structure of nonionic surfactant aggregates formed by self-assembly at the surface of spherical silica particles of 8.2 nm radius. SANS scattering profiles from dilute dispersions of this silica in contrast-matching $H_2O/D_2O$ containing a concentration of surfactant two orders of magnitude higher than the *CMC* show that the maltoside surfactant $\beta$-$C_{12}G_2$ interacts very weakly with the silica particles and the scattering profiles resemble those of free micelles which have an oblate ellipsoidal shape. This result conforms with earlier findings of low adsorption levels of sugar surfactants at macroscopic silica surfaces,[18,33] and shows that the adsorption of these surfactants is not enhanced by strong surface curvature. On the other hand, the alkyl ethoxylate surfactant $C_{12}E_5$ exhibits strong cooperative adsorption onto the silica particles with a surface concentration plateau value $\Gamma_{mx}$ similar to that on flat silica surfaces. The SANS profiles from the dilute dispersion in contrast-matching $H_2O$/$D_2O$ exhibit a local maximum at intermediate $q$ values, which becomes more pronounced with increasing surface concentration of the surfactant. A core-shell model corresponding to a uniform surfactant layer at the surface of the silica perticles accounts well for the increase in film thickness with increasing surface concentration, but this model underestimates the surface area of the adsorbed surfactant as extracted from the high-$q$ regime of the scattering spectra. The SANS spectra can be represented very well by the model of micelle-decorated silica beads on the assumption that $C_{12}E_5$ is adsorbed as individual surface aggregates of spherical geometry. According to this analysis an increase of the surface concentration of the surfactant leads to an increasing number of surface aggregates, up to a close packing of spherical surface micelles. The preference of such small surface aggregates is attributed to the high surface curvature of the silica nanoparticles which prevents an effective packing of the hydrophobic chains in a bilayer configuration.

**Acknowledgments**

The authors wish to thank Dirk Berger for help with the Transmission Electron Microscopy, and one of the Referees for drawing attention to an error in the adsorption isotherm. D.L. is grateful to Deutscher Akademischer Austauschdienst (DAAD) and to the Fundación Gran Mariscal de Ayacucho (Fundayacucho) for receiving a doctoral scholarship. Financial support by DFG through project FI 235/15-2 and the cooperation initiated in the framework of the French-German Network "Complex Fluids: From 3 to 2 Dimensions" (Project FI 235/14) is also gratefully acknowledged.

**Appendix: SANS Analysis**

For a monodisperse system of spherically symmetric silica particles in a solvent, the coherent scattering cross section can be modeled by the following expression:[34]

$$I(q) = \phi \Delta\rho^2 V_b F(q) S(q), \tag{A1}$$

where $\phi$ denotes the volume fraction of the particles, $\Delta\rho$ is the difference in scattering length density between the silica particles and the solvent/matrix, $V_b$ is the volume of a particle. The structure factor $S(q)$ describes the spatial correlations between particle positions arising from the repulsive (excluded volume, electrostatic) or attractive potential between the particles. For sufficiently dilute systems there is no position-position correlation and $S(q) = 1$. The form factor $F(q)$ depends on the shape of the particle and the scattering length distribution within the particle. For a particle with spherical geometry it is given by [35]

$$F(q) = |f(qR)|^2$$

with

$$f(x) = 3 \cdot \frac{\sin(x) - x\cos(x)}{x^3} \tag{A2}$$

Here, $R$ is the radius of a particle, and $x = q \cdot R$, where $q$ is the magnitude of the scattering vector. The polydispersity in size of the silica particles is described by means of a log-normal distribution with parameters $R_{Si}$ and polydispersity $\sigma$.

$$P(R, R_{Si}, \sigma) = \frac{N/V}{\sqrt{2\pi}R\sigma} \exp\left(-\frac{1}{2\sigma^2}\ln^2\frac{R}{R_{Si}}\right) \tag{A3}$$

where $N/V$ is the number density of silica particles.

The form factor with polydispersity is calculated by integration:

$$F(q) = \int P(R, R_{Si}, \sigma) F(q, R) dR \tag{A4}$$

The scattering from the shell of thickness $\delta$ and internal radius $R$ can be described by: [6,8,32,36]

$$I(q) = \frac{N}{V}\Delta\rho^2 \left[\frac{4\pi}{3}(R+\delta)^3 f(q(R+\delta)) - \frac{4\pi}{3}R^3 f(qR)\right]^2 \quad (A5)$$

$f(x)$ is defined in eq. (A2); the polydispersity of the radius of silica particles could be included by integration similarly as in eq. (A4).

For a spherical silica particle with adsorbed surfactant as isolated, identical and spherical surface aggregates, which interact only through excluded volume interactions, the scattered intensities can be calculated by:[8]

$$I(q) = \frac{N}{V} N_{sph} \Delta\rho^2 S(q) S_{mic}(q) F_{mic}^2(q) V_{mic}^2 \quad (A6)$$

where the form factor $F_{mic}^2$ is the normalized form factor of free micelles of volume $V_{mic}$, modeled as monodisperse spheres; $S(q)$ is the inter-aggregate structure factor and $S_{mic}(q)$ is the inter-micellar structure factor of micelles sticking to one silica particle, (cf. ref. 8 for details). In the calculations of $S_{mic}(q)$ the polydispersity in the radius of the silica particles, which leads to a polydispersity in the number of adsorbed micelles, is taken into account. We have convoluted the data with the resolution function. The finite resolution of the instruments has been taken into account in the theorethical fitting functions as is shown in ref. 8.